# Response of Seven Crystallographic Orientations of Sapphire Crystals to Shock Stresses of 16 to 86 GPa

G. I. Kanel,[1] W. J. Nellis,[2] A.S. Savinykh,[3] S. V. Razorenov,[3] and A. M. Rajendran[4]

[1]Joint Institute for High Temperatures, Moscow, 125412 Russia
[2]Department of Physics, Harvard University, Cambridge, MA 02138, USA
[3]Institute of Problems of Chemical Physics, Chernogolovka, 142432 Russia
[4]U.S. Army Research Office, RTP, NC 27709-2211

Abstract

Shock-wave profiles of sapphire (single-crystal $Al_2O_3$) with seven crystallographic orientations (*c*, *d*, *r*, *n*, *s*, *g*, and *m*-cut) were measured with time-resolved VISAR interferometry at shock stresses in the range 16 to 86 GPa. Shock propagation was in the direction normal to the surface of each cut. The angle between the *c*-axis of the hexagonal representation of the sapphire crystal structure and the direction of shock propagation varied from 0 for *c*-cut up to 90 degrees for *m*-cut in the basal plane. Based on published shock-induced transparencies for 3 directions of shock propagation, shock-induced optical transparency correlates with the smoothness of the mechanical shock-wave profile. The ultimate goal was to find the direction of shock propagation for which shock-compressed sapphire is most transparent as a window material. In the experiments particle velocity histories were recorded at the interface between a sapphire crystal and a LiF window. In most cases measured wave profiles are noisy as a result of heterogeneity of deformation. Measured values of Hugoniot Elastic Limits (HELs) depend on direction of shock compression and peak shock stress. The largest HEL values (24 GPa) were recorded for shock loading along the *c*-axis and perpendicular to *c* along the *m*-direction. Shock compression along the *m*- and *s*-directions is accompanied by the smallest heterogeneity of deformation and the smallest rise time of the plastic shock wave. *m*- and *s*-cut sapphire most closely approach ideal elastic-plastic flow, which suggests that *m*- and *s*-cut sapphire are probably the sapphire orientations that remains the most transparent to the highest shock pressures. Under purely elastic deformation sapphire demonstrates very high spall strength, which depends on both load duration and peak stress. Plastic deformation of sapphire causes loss of its tensile strength.

# I. Introduction

Strong dielectric materials that remain transparent under shock compression have been sought for decades for scientific and technological applications. Polycrystalline ceramics, such as alumina (ceramic $Al_2O_3$), are not transparent at ambient and are opaque under shock compression, as well. In contrast, single crystals of many dielectrics are transparent at ambient and, thus, are transparent in some range of shock pressures. However, strong dielectric crystals are yet to be found that are transparent over a sufficiently wide range of shock pressures to be useful for the wide range of possible applications.

When shock-wave experiments are performed on traditional crystals, such as quartz (single-crystal $SiO_2$) and sapphire (single-crystal $Al_2O_3$), the crystals light up optically and heterogeneously at shock pressures comparable to material strength [1]. In addition, the mechanical response has large data scatter, which suggests that deformation is heterogeneous and statistical on a length scale of microns or more [2]. Sapphire, for example, is commonly used in shock-compression experiments as anvils, tampers, and windows. Anvils are used to double-shock or multiply shock a sample material with lower shock impedance than the anvil. A tamper is used to prevent a sample with higher shock impedance than the tamper from releasing down to zero pressure when a shock front reaches the free surface of that sample. If the anvil or tamper is transparent, then temperature can be determined from thermal radiation emitted from the shocked sample or a shocked sample can be probed with an incident laser beam.

A prime example of the latter is a VISAR (Velocity Interferometer for a Surface of Any Reflector), which measures the history of material velocity induced by a shock wave. Such histories are commonly known as shock-wave profiles and are measured to detect elastic-plastic response and phase transitions. In this method a shock is generated on the front surface of a sample and a transparent crystal is placed flush against its rear surface to tamp that surface when the shock reaches the rear surface of the sample. A laser beam is transmitted through the transparent crystal anvil/window to reflect off the rear surface of the sample. The history of the Doppler shift of the reflected laser beam is a measure of the velocity history of the interface between sample and window. The first such experiment was performed by Barker and Hollenbach, who used a number of windows, including *c*-cut sapphire [3]. A *c*-cut sapphire disk (0001) is one that is cut so the normal to a flat surface of the disk is parallel to the *c* axis of the hexagonal representation of the crystal structure. Barker and Hollenbach found that ~15 GPa (150 kbar) is the highest shock pressure at which good VISAR data was obtained with sapphire. At higher pressures "shock-induced luminescence and/or loss of transparency" of sapphire were observed.

Graham and Brooks then measured the Hugoniots of *c*-cut, *a*-cut (11-20), and *n*-cut (11-23) sapphire crystals [4]. The normals to the flat surfaces of *a*-cut and of *n*-cut crystals are parallel to the *a* axis and *n* axis, respectively. The *a* axis is in the basal plane of the hexagonal crystal structure. The *n* axis makes an angle of $61^0$ with the *c* axis. The Hugoniot elastic limits (HELs) varied from 21 GPa for shock propagation along the *c* direction to 12 GPa for shock propagation along the *n* direction. Above the HELs (~15 GPa) the Hugoniots of all three orientations are essentially identical, which was interpreted as caused by a substantial loss of shear strength. The shear-stress offset of the Hugoniot above the HELs was estimated to be about 5 GPa relative to the 300-K isotherm. It was suggested that the decrease in shear strength might be caused by substantial shock-induced microstructural damage. While those data were analyzed under the assumption of steady wave structures, the measurements above ~15 GPa were done using a rotating-mirror streak camera, which is weakly sensitive to time dependences of non-steady shock waves. Based on shock propagation in three different crystallographic directions, those results suggest that the sapphire Hugoniot is isotropic above the HELs and that *c*-cut sapphire is unreliable in VISAR experiments at shock pressures above 15 GPa because of microstructural damage.

The apparently isotropic nature of the Hugoniot of hexagonal sapphire has implicitly been interpreted to mean that the optical response of sapphire on its Hugoniot above 15 GPa is also isotropic. For this reason the great majority of shock experiments that have used sapphire as a window have used *c*-cut crystals, which are readily available commercially. As shown below shock-induced opacity on the Hugoniot of sapphire is certainly not isotropic. Further, Graham and Brooks interpreted there Hugoniot data under the assumption that their shock waves in sapphire were steady. As shown below shock waves are not steady in sapphire above ~10 GPa. Further, the wave profiles below indicate that the concept of Hugoniot Elastic Limit (HEL) is essentially without meaning for sapphire single crystals shocked above ~10 GPa. The reason is that the elastic wave in sapphire starts to relax after only a few nanoseconds at peak elastic pressure. While HEL is defined mathematically to be the peak elastic stress, it has little physical significance in sapphire because the lifetime of the peak may be only a few ns.

Hyun et al measured the HELs of *c*-cut and *a*-cut single crystals to be greater than 12.5 GPa and the HEL of *r*-cut (1-102) to be 6.5 GPa [5]. The *r* direction is $57^0$ from the *c* axis, close to that of the *n* direction, $61^0$, investigated by Graham and Brooks. The velocity of sound at ambient of single-crystal $Al_2O_3$ was measured as $11.23 \pm 0.05$ km/s along the *a* and *c* directions and $10.47 \pm 0.04$ along the *r* direction [6].

Ahrens et al measured the shock compression curve of polycrystalline ceramic $Al_2O_3$ (Lucalox) and found an HEL of 11 GPa and a shear offset on the Hugoniot of 3-4 GPa up to 30 GPa relative to the 300-K isotherm. The calculated temperature rise from the initial temperature of 300 K is ~100 K at a shock stress of 30 GPa. The corresponding thermal pressure is ~0.1 GPa, negligible compared to the shear strength [7]. Mashimo et al measured the Hugoniot of $Al_2O_3$ along the *a* axis up to 110 GPa, found scatter in the HEL, and reported a phase transition at 79 GPa [8]. Using shock–reshock loading techniques, Reinhart et al [9] did not confirm the transition but did observe that single-crystal sapphire has considerable loss of strength at approximately 56 GPa on the Hugoniot.

Munson and Lawrence [10] and Grady [2] measured shock-wave profiles in ceramic $Al_2O_3$ in the pressure range 10 to 40 GPa using a VISAR [2]. Those profiles are elastic-plastic with a rounded transition between the two. They are essentially what Graham and Brooks expected them to be for their single crystals. Our shock-wave profiles for sapphire single crystals (shown below) are substantially different from what Grady measured for ceramic samples. As for Graham and Brooks [4], Grady's work on ceramic samples is excellent, but the results are not representative of shock-wave profiles in single-crystal $Al_2O_3$, as they have been expected to be.

Dandekar et al [11] measured shock-wave profiles in ceramic $Al_2O_3$ (AD995) and found that the longitudinal shock waves rise sharply in samples shocked to 3.9 and 6.9 GPa, while the rise times are 0.1 to 0.2 µsec for samples shocked to 8.4 and 11.1 GPa. The HEL is 6.7 GPa and the shear stress offset increases up to a maximum of 2 to 4 GPa at a shock stress of 20 GPa and then decreases to zero at 30 GPa. They also measured transverse profiles, which have complex shapes. Zaretsky and Kanel [12] performed experiments with controlled variation of the transverse stress in alumina ceramic specimens subjected to planar impact, which demonstrated ductile response of the ceramic under conditions of one-dimensional shock compression.

Shock-induced opacities of sapphire have been investigated. Urtiew observed that shock-compressed sapphire loses some of its transparency in the infrared at stresses in the range 100 to 130 GPa [13]. More recently, the electrical resistivity of shocked sapphire was found to begin to decrease significantly with increasing shock pressures above 130 GPa [14], which is consistent with Urtiew's results.

At shock pressures between 16 and 85 GPa, Kondo found that optical spectra emitted from shock-compressed sapphire could be described as gray-body with an effective temperature of $5600 \pm 500$ K and an emissivity that increases continuously up to 0.08 at the highest shock pressure [15]. These temperatures are larger by an order of magnitude than those estimated by simple thermodynamic considerations [7]. These results were interpreted as heterogeneous emission from hot spots whose areal density increases continuously with shock pressure, thus causing the emissivity to increase continuously as the emitting area increases with shock pressure [15].

Kwiatkowski and Gupta studied optical spectra in the range 450 to 650 nm of *c*-cut sapphire shocked above the HEL [16]. They observed weak, broadband emission spectra, which is too narrow to be gray-body. That is, at shorter wavelengths the emitted intensity is reduced relative to a gray body. At least part of the extinction is probably caused by scattering because the spectral dependence of the extinction is consistent with a distribution of Mie scatterers that are small compared to a wavelength of light.

Hare et al [17] took fast-framing photographs of $Al_2O_3$ single crystals as they were shock-compressed along the *a*, *c*, and *r* directions. The *r* direction makes an angle of $57^0$ with the *c* axis. They observed that *r*-cut crystals are essentially optically transparent up to 47 GPa while *a*-cut and *c*-cut crystals are opaque to a significant degree well below 47 GPa. The implication derived from these experiments is significant. While the Hugoniot is probably isotropic above 20 GPa [4], optical transparency under shock is not. Further, to obtain maximum optical transparency under shock compression, the choice of sapphire crystal orientation is important. The optimum orientation is somewhere between the principal directions, that is, along *c* and in the basal plane. To date optical photographs of sapphire have only been made on crystals oriented with shock propagation in one direction between the *c* axis and the basal plane – along the *r* axis.

Shock recovery experiments have been performed on sapphire samples by Wang and Mikkola [18]. They found that plastic deformation via slip and twinning [19, 20] is sensitive to the direction of shock propagation in the sapphire lattice, being largest in the *r* direction relative to the *a* and *c* directions. When compared to the observation that optical transparency under shock is greatest for shock waves traveling in the *r* direction, plastic deformation is apparently the desirable deformation mechanism to retain maximum optical transparency under shock compression. Basal twins and partial dislocations in basal planes are two of the most commonly observed microstructures in shock-wave-deformed sapphire. Chen and Howitt [21] have found that the partial dislocations are the twinning dislocation for basal twins. A large volume of information about deformation mechanisms, including basal and pyramidal slips and basal and rhombohedral twins, in sapphire under hypervelocity impact has been obtained by means of molecular dynamic simulations [22].

A major unresolved issue is identification of an initially strong crystal that is optically transparent over a substantial range of shock pressures and determination of the mechanism that induces opacity under shock. In this paper we report measurements of shock-wave profiles of sapphire single-crystals. Sapphire was chosen for experimental investigations because of its high strength, relatively high shock impedance (close to that of iron, for example), and the body of experimental data that exists to guide the design of experiments and the interpretation of results. Especially important in this regard is the optical imagining experiments on sapphire crystals oriented such that shock waves were traveling along the *c, r,* and *a* directions [17]. That work indicates that single crystals should be investigated because optical emission is sensitive to the direction of shock propagation in the sapphire crystal lattice.

Our fundamental assumption in our search for maximum transparency under shock is the expectation that there is a correlation between mechanical response measured by shock-wave profiles and optical response measured by fast-imaging photographs. We assume further that heterogeneous mechanical deformation and the degree of shock-induced light are coupled and that shock-induced opacity is strongly coupled to both. Our working hypothesis is to find directions of shock propagation for which shock-wave profiles are smoothest; that is, data scatter in shock-wave profiles is least, because optical transparency is expected to be greatest in those directions. We therefore expect that by minimizing data scatter in shock-wave profiles, optical transparency under shock will be maximized. In this paper we report measurements of shock-wave profiles measured for sapphire single crystals of seven orientations at nominal shock stresses of 16 GPa, 22 – 24 GPa, and 86 GPa, which are (i) near the Hugoniot elastic limit, (ii) in the so-called "elastic-plastic range", and (iii) close to overdriving the first (elastic) wave, respectively.

**II. Experiments**

Longitudinal sound speed, $c_l$, of each crystal orientation was measured to look for a possible correlation between $c_l$ and the nature of crystal response to longitudinal shock propagation. The samples were high-quality sapphire discs of seven different orientations 2.4 or 5.05 mm thick and 22 or 50 mm in diameter, respectively. Hexagonal crystal orientations in this study are indicated by 4 indices, 3 in the basal plane and one along the *c* axis. The orientations investigated are listed in Table 1 with the angle $\Theta$ subtended between the normal to each disk and the *c* axis.

**Table 1**. Crystal orientations of sample disks in this study. Directions of shock propagation are perpendicular to flat surfaces of disks, which are labeled with 4 hexagonal indices; $\Theta$ is angle between normal to each disk and *c* axis. $c_l$ is longitudinal sound speed.

| Orientation | 4 indices | $\Theta$, ° | $c_l$, km/s |
|---|---|---|---|
| *c*-cut | (0001) | 0 | 11.19 |
| *d*-cut | (10$\bar{1}$4) | 38 | 10.95 |
| *r*-cut | (1$\bar{1}$02) | 57 | 10.60 |
| n-cut | (11$\bar{2}$3) | 61 | 10.85 |
| s-cut | (10$\bar{1}$1) | 72 | 11.20 |
| g-cut | (11$\bar{2}$1) | 79 | 11.11 |
| m-cut | (10$\bar{1}$0) | 90 | 11.21 |

Figure 1 presents schematically the experimental arrangement. Planar-compression waves were generated in sapphire samples mounted on 2.0 mm-thick base plates by impacts of 2.0 mm-thick aluminum flyer plates. The flyer plates were launched using explosive facilities. Impact velocities of Al plates onto Al base plates backed by sapphire crystals were 1.2±0.05, 1.8±0.05, and 5.2±0.1 km/s. With these impact velocities peak stresses in shock-compressed sapphire were around 16 GPa (elastic compression), 23 GPa and 86 GPa, correspondingly. Particle velocity histories were recorded with a VISAR [23] at the interface between the sample and LiF "window". We used either vacuum-sprayed aluminum coating or aluminum foils 7 μm in thickness as a reflector for the VISAR beam. In most experiments the velocity-per-fringe constant of the VISAR was $u_0$ = 238.1 m/s, which took into account the window correction [3]. Some of the experiments were done with $u_0$ = 419.1 m/s in order to justify the absolute velocity values behind the elastic and plastic shock waves. In addition, absolute velocity values were estimated by means of computer simulations of performed experiments with simplified models of the materials.

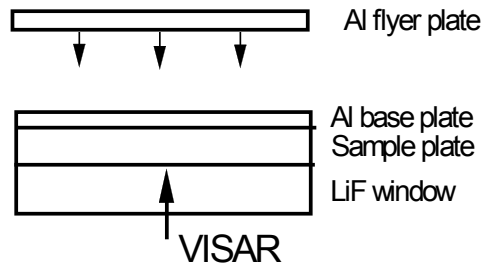

Fig. 1. Schematic of experimental arrangement for recording waveforms at interface between sapphire sample and LiF window.

**III. Results of measurements**

  **III.1. *c*-cut sapphire**

Figure 2 presents the VISAR waveforms measured for 5.0 mm-thick *c*-cut sapphire samples at intermediate peak stress. Also shown are results of computer simulations used to verify absolute values of measured velocities and in this way to estimate the number of "lost fringes" in the VISAR interferograms. The waveforms for sapphire differ qualitatively from those recorded for fine-grain alumina ceramic. The particle velocity histories in Fig. 2 are "noisy" and their reproducibility is much less than the case of alumina ceramic [20, 22] whose waveforms are smooth and well reproducible. The strong irregular oscillations and low reproducibility of the waveforms of *c*-cut sapphire are a consequence of intrinsic heterogeneity of its inelastic deformation. The general shapes of the waveforms are also different. In the case of alumina ceramic particle velocity increases monotonically between the elastic precursor wave and the plastic compression wave, whereas for sapphire we see a spike behind the elastic wave front.

The rise time of the precursor wave front does not exceed 2 ns. The rise time of the second compression wave is around 150 ns. The estimated average propagation speed of the midpoint of the second compression wave is 7.4 km/s for the 5-mm sample, which is lower than the bulk sound speed value 8.0 km/s evaluated from the value of 255 GPa of the sapphire bulk modulus. It has been shown [24] that the precursor waveforms with spikes are associated with accelerating stress relaxation behind the precursor front. Although the wave, in general, is compressive, in this case stress relaxation occurs so rapidly that it causes rarefaction behind the elastic front. The elastic spike has a transient nature and may only appear during the process leading to establishment of the wave. Computer simulations [24] have also demonstrated subsonic speed of the plastic compression wave in the initial phase in this case.

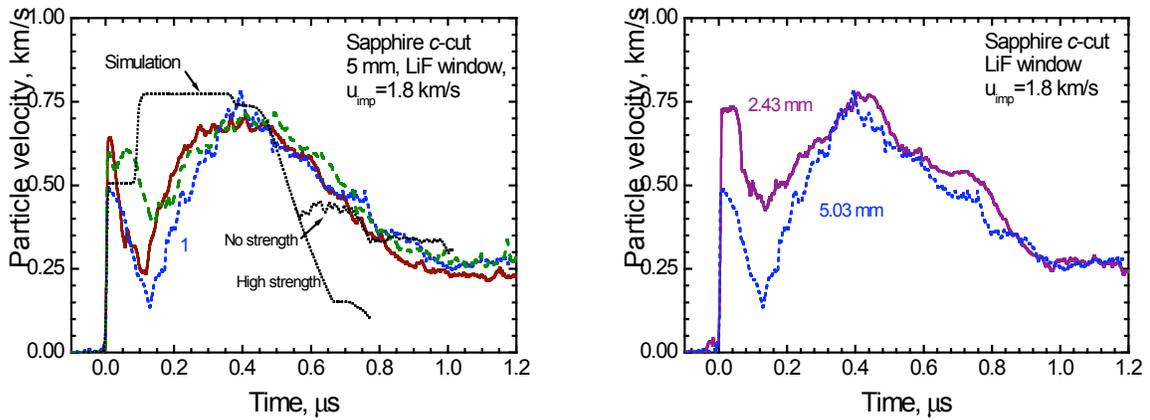

Fig. 2. Results of three experiments with 5 mm-thick *c*-cut sapphire samples impacted by 2 mm-thick aluminum flyer plates at 1.8±0.05 km/s onto a 2 mm-thick aluminum base plate backed by sapphire. Particle velocity histories were recorded at the interface between the sapphire sample and the LiF window.

Fig. 3. Comparison of the particle velocity history for *c*-cut sapphire 5.03 mm in thickness (waveform 1 in Fig. 2) with the waveform measured for 2.43 mm-thick sample impacted by aluminum flyer plate 2 mm-thick at 1.8±0.05 km/s velocity.

Also shown in Fig. 2 is a computer simulation of the shock response of a *c*-cut crystal assuming an idealized elastic-plastic strength model on compression and assuming both a high and zero tensile strength on release of pressure. The reasonable agreement between the measured velocities on release at late times above 0.5 µs and the simulation with zero tensile strength indicates negligible tensile strength after deformation in the plastic wave. It is interesting to note that after taking into account the different shock impedances of sapphire and the LiF window that the deep valley behind the elastic spike is only observed if the spall strength of *c*-cut sapphire in the state between the elastic and plastic waves is 2–2.5 GPa or larger. Since sapphire looses its strength after inelastic deformation, the spike-like precursor wave is obviously associated with purely elastic deformation in both compression and unloading and is not associated with compressive fracture.

Figure 3 shows results of measurements at two different sample thicknesses. The data demonstrate strong decay of the HEL with increasing propagation distance. It can be seen the time interval between the elastic precursor front and the second wave is practically the same for these two propagation distances. Perhaps such unusual response denotes a delayed nucleation of plastic shears in sapphire, which does not necessarily occur on its impact surface. Note that 150-200 ns of the rise time of the second wave correspond to 1–1.5 mm of wave-propagation distance, which is close to the 2.43 mm thickness of the thinner sample. In other words, the experiment with thin sample gives us information about establishing the waves rather than about evolution of steady waves.

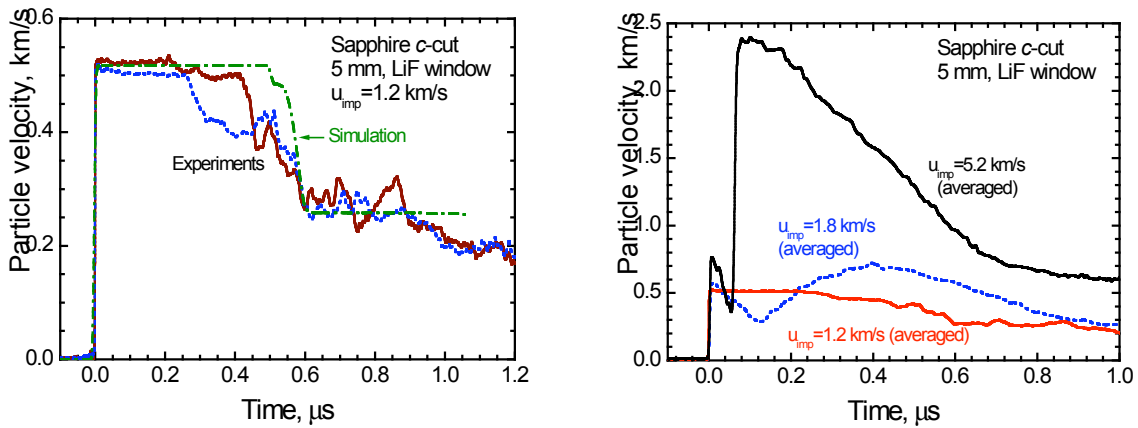

Fig. 4. Results of two experiments with 5 mm-thick *c*-orientated sapphire samples impacted by an aluminum flyer plate at 1.2±0.05 km/s and the result of a computer simulation of the experiment assuming a perfectly elastic response of sapphire.

Fig. 5. Averaged particle velocity histories of 5 mm-thick *c*-cut sapphire samples at three different impact stresses.

Figure 4 presents results of two shots with *c*-cut sapphire at low impact velocity (1.2±0.05 km/s). Comparison with a computer simulation confirms initially purely elastic response and stress relaxation after approximately 0.25 μs of delay time at 16 GPa peak stress. Until the moment of the beginning of stress relaxation, the waveforms are smooth, and irregular oscillations appear simultaneously with the stress relaxation. Thus, it appears that plastic shears nucleate in a random manner.

Figure 5 summarizes experimental data for *c*-cut sapphire samples 5 mm in thickness at three different impact stresses. The waveforms shown are averaged data of two shots at an impact velocity of 1.2 km/s, three shots at 1.8 km/s and two shots at 5.2 km/s. The particle velocity histories measured at highest peak stress are less noisy compared to those recorded at lower impact stresses and are more reproducible. The rise time of the second compression wave is about 5 ns. These particle-velocity histories demonstrate strong dependences of the recorded HELs and of the rate of stress relaxation on peak stress in the plastic shock wave. Higher stress at the elastic precursor front is accompanied with faster stress relaxation behind it. An important aspect of the leading parts of the waveforms is the intersection between elastic and plastic fronts: at higher peak value of the stress its relaxation each time occurs deeper than at lower peak stress. Such evolution of the waveforms is unusual; as a rule waveforms vary monotonically with increasing peak stress. It is likely that, as a result of hydrodynamic decay of the spike-like precursor wave, HEL values at different peak stresses for 5-mm-thick samples will vary not only in value but even in their sequence with increasing propagation distance.

HEL values were determined from the recorded waveforms by means of analysis the wave interactions at the sapphire-LiF interface. The aim of the analysis was to determine the elastic precursor wave parameters in sapphire using the measured values of the particle velocity $u_{HEL}$ at the boundary between a sapphire sample and a lithium fluoride window. The average HEL of $c$-cut sapphire at the intermediate peak stress is 17.7 GPa with a scatter from 14.9 to 19.9 GPa.

Experiments on $c$-cut sapphire in the region of its HEL indicate a delayed beginning of inelastic deformation, which is accompanied by a loss of tensile strength. In order to verify elastic behavior of sapphire during the delay time we performed additional experiments aimed at determining spall strength at various peak stresses and load durations. Figure 6 shows particle velocity histories recorded at the interface between a 5 mm-thick $c$-cut sample and a water window, rather than LiF. The samples were placed on 2 mm-thick aluminum base plates, which were impacted by aluminum flyer plates 0.4 mm or 0.85 mm thick at 1.55 km/s.

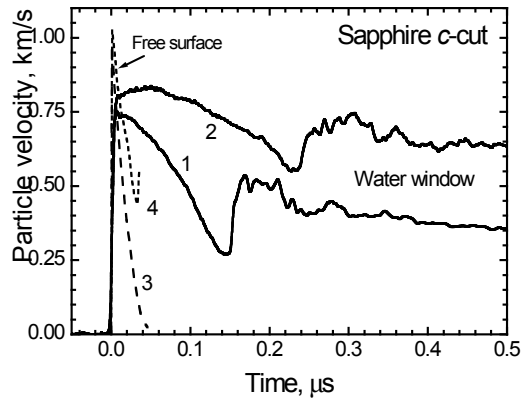

Fig. 6. Results of experiments on measurement of sapphire spall strength. 1, 2 are the velocity histories of interfaces between water windows and 5-mm thick $c$-cut samples impacted by flyer plates 0.4 mm (curve 1) or 0.85 mm (curve 2) in thickness at 1.55 km/s. 3, 4 are free surface velocity histories of sapphire samples from ref. [25].

Tensile stresses are generated inside a sapphire sample after reflection of the compression pulse from the low-impedance water window. When the peak tensile stress reaches a sufficiently high value, spall fracture is initiated inside the sample. As fracture develops, the tensile stresses relax to zero. As a result, a compressive disturbance called a "spall pulse" appears in the interface velocity profile as a second velocity rise. It can be see that the waveforms are smooth and are not distorted by any oscillations in their initial parts, but the irregular oscillations appear immediately when fracture begins. Since it follows from discussion above that inelastic deformation of $c$-cut sapphire is accompanied with significant perturbations of the stress field, absence of the later may be considered as evidence of elastic response during shock compression and subsequent tension up to the beginning of fracture.

Stress values immediately before fracture (spall strengths) were calculated by the characteristic method using measured peak velocity value and minimum particle velocity ahead of spall pulse and the Hugoniots of sapphire and water [26]. In the shot with the thin flyer plate, some decay of the shock wave occurred as a result of which the peak compressive stress near the sample-window interface was 18.2 GPa. In this shot sapphire demonstrated a spall strength as high as 8.9 GPa, which is larger than the spall strength of any armor steels. In shot 2 with a larger pulse duration, the peak compressive stress was 20.6 GPa and the spall strength was 4.2 GPa. Figure 6 also shows the free surface velocity histories of *c*-cut sapphire samples measured earlier [25] with 40-60-ns load durations. Spall was not recorded on shot 3, where the peak stress in elastic compression wave was 23 GPa and the tensile stress in the reflected rarefaction wave reached 20 GPa. An increase of compressive stress up to 24 GPa in shot 4 and some increase of the total load duration resulted in spallation at a factor of two less tensile stress – 10.4 GPa. Thus, at shock compression below the HEL, sapphire demonstrates the highest values of spall strength, which grow with shortening of the load duration. There is also a trend for the spall strength to decrease with increasing peak compressive stress in the range of the HEL.

### III.2. *r*-cut sapphire

Figures 7 to 9 summarize results of experiments with *r*-cut sapphire samples at three different peak stresses. With the exception of the lowest impact stress (Fig. 8), these waveforms are less oscillated and show lower HEL value than observed for *c*-cut sapphire. In Fig. 8 the recorded step-like unloading in 2 mm-thick sample plate is a result of multiple wave reflections in sapphire between the LiF window and the Al base plate. Also at the smallest impact velocity the waveforms are less reproducible in their initial parts, where they show a relaxation from the state corresponding to purely elastic response in one shot and ramped growth of the stress behind the elastic wave front in other shot. The rise time of the plastic compression wave in *r*-cut sapphire at intermediate peak stress is around 85-90 ns, which is much less than the rise time in *c*-cut sapphire. At lowest impact velocity the waveforms are noisy and their reproducibility is lower, which may be considered as evidence of random nucleation of plastic shears.

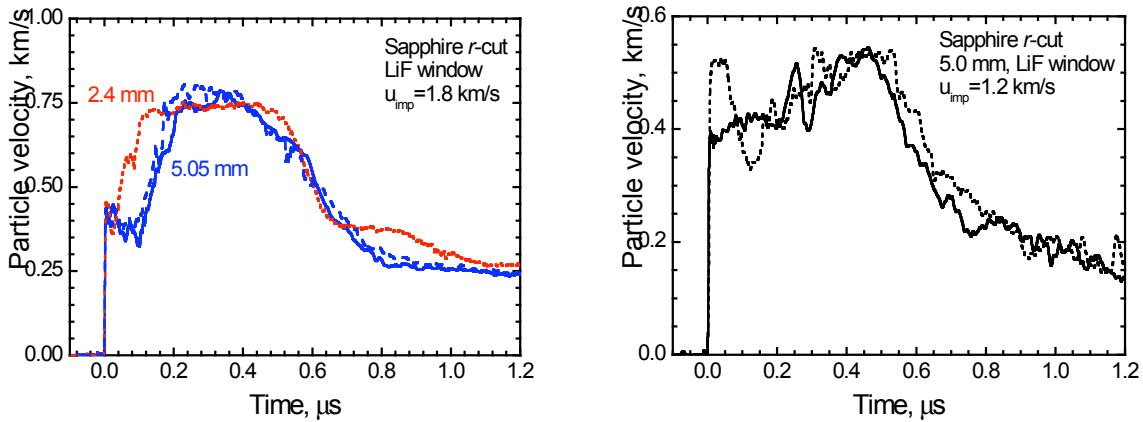

Fig. 7. Results of experiments with *r*-cut sapphire samples 5.05 mm (two shots) and 2.4 mm thick impacted by a 2-mm aluminum flyer plate at 1.8 km/s. The aluminum base plates were 2 mm thick in all three cases.

Fig. 8. Results of two experiments with 5 mm-thick *r*-cut sapphire samples.

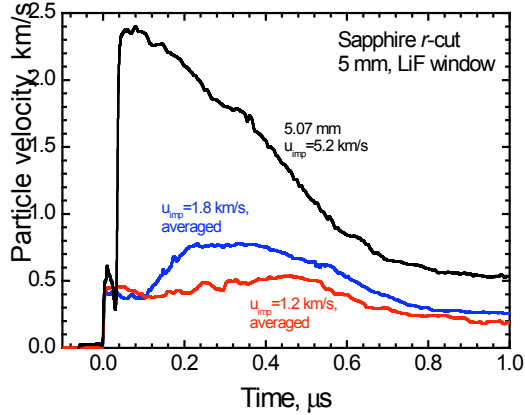

Fig. 9. Results of experiments with 5-mm thick *r*-cut sapphire samples at three different impact velocities.

Duvall [27] has proposed a theory describing the attenuation of elastic precursor waves. The observed decrease in the precursor stress with propagation path length is accounted for by assuming that the shocked material will momentarily support a value of shear stress higher than equilibrium. As the shock front passes, the shear stress relaxes toward its equilibrium value and caused release reduces the stress behind the precursor front. At higher impact stress the precursor decay process starts from higher initial stress and approaches equilibrium value after larger propagation distance. As a result HEL values measured at various impact stresses at the same sample thickness are usually larger at larger impact stress or do not remarkably depend on the impact stress. Comparison of the waveforms in Fig. 9, however, shows an abnormal dependence of HEL on peak stress. Rather than the usual observation of monotonic growth, the stress at the front of the elastic precursor at intermediate impact velocity is less than for a lower impact velocity.

### III.3. *s*-cut sapphire

Figures 10 and 11 summarize results of experiments with 5 mm-thick *s*-cut sapphire samples at three different peak stresses. At intermediate stress, sapphire of this orientation demonstrates high reproducibility of the waveforms and a relatively small rise time (30 ns or less) of the plastic wave, which is much less than the rise times of *c*-cut and *r*-cut sapphire. There still are some irregular oscillations in the waveforms but they are of higher frequencies and lower amplitudes than those recorded in experiments with other orientations. The speed of second (plastic) wave is 8.3 km/s, which is in reasonable agreement with the bulk modulus of sapphire.

The step-like structure of the recorded plastic wave is a result of multiple reflections of the elastic-precursor wave. Reflection of the elastic precursor from the sapphire/LiF interface causes elastic unloading from the HEL of 13.1 GPa down to 6.5 GPa in accordance with the shock-impedance mismatch of sapphire and LiF and the appearance of a reflected elastic unloading wave. When the reflected unloading wave meets the plastic shock, a new elastic compression wave is formed in the unloaded layer. This can be interpreted as a reflection of the elastic wave at the plastic shock front. This reflection forms an intermediate step in the recorded velocity profile of the plastic wave at about 160 ns in Fig. 10. The time interval between this intermediate step and the main plastic wave, as well as the stress increment in the intermediate step are in agreement with this explanation.

Accounting for the distortion of the plastic wave by multiple reflections of elastic waves its real rise time is estimated as measured rise time minus the calculated time interval between the reflected elastic wave and the plastic wave that gives 11–12 ns. A small rise time and high-frequency "noise" may be considered as evidence of high homogeneity of deformation of *s*-cut sapphire under shock compression. As for most of the orientations studied, the shot at an impact velocity of 1.2 km/s shows a higher HEL than that at 1.8 km/s and stress relaxation after 0.12 μs delay time. The plastic shock wave at an impact velocity of 5.2 km/s contains an additional step as well as that at 1.8 km/s. Accounting for the multiple reflections of the elastic wave, the rise time of the plastic shock wave at 5.2 km/s is evaluated from the measurements as 5-7 ns. Figure 11 shows that two experiments on *s*-cut sapphire at an impact velocity of 1.8 km/s essentially reproduce each other.

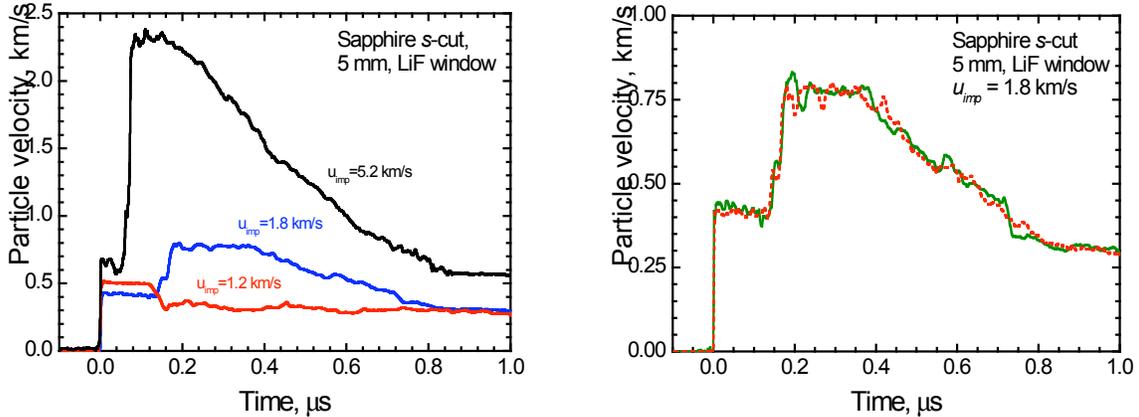

Fig. 10. Results of experiments with s-cut sapphire samples at three pressures.

Fig. 11. Results of two experiments with 5 mm-thick *s*-cut sapphire samples on 2 mm-thick aluminum base plates with 2 mm-thick aluminum flyer plates at an impact velocity of 1.8 km/s.

### III.4. *m*-cut sapphire

Figures 12 and 13 present results of experiments with *m*-cut sapphire samples at different peak stresses. For comparison, Fig. 13 shows the data for *m*-, *s*- and *c*-cut sapphire. The data of *m*-cut sapphires are very reproducible and much less noisy than the waveforms of most other orientations. The rise time of the plastic wave at the intermediate peak stress is about 5 ns, the smallest one measured at the intermediate pressure. The HEL of *m*-cut sapphire is close to the HEL of *c*-cut samples but, unlike the latter, stress relaxation behind the elastic-precursor front is small at intermediate and highest peak stresses. The HEL increases monotonically with peak stress and the waveforms in Fig. 12 never intersect each other. Of all the cuts investigated, *m*-cut sapphire most closely approaches ideal elastic-plastic flow.

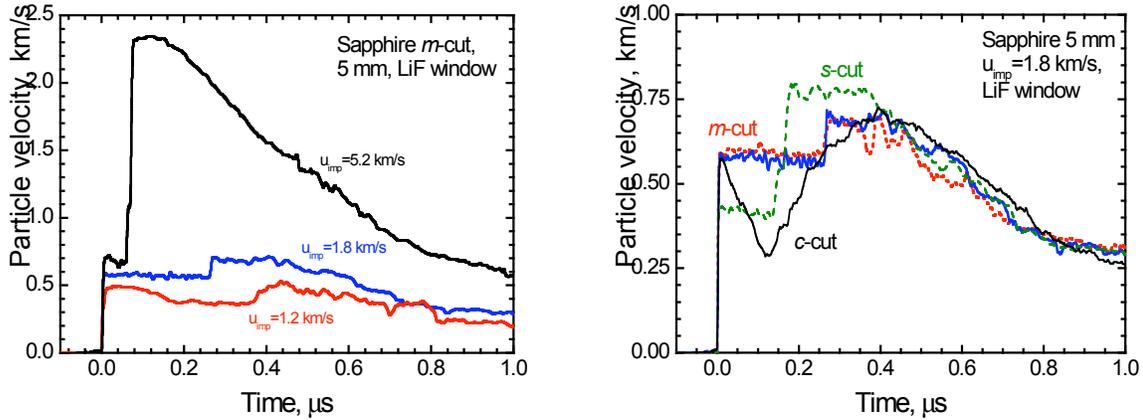

Fig. 12. Results of experiments with *m*-cut sapphire samples at three different peak stresses.

Fig. 13. Results of two experiments with 5 mm-thick *m*-cut samples on 2 mm-thick base plates impacted by 2 mm-thick aluminum flyer plates at 1.8 km/s. For comparison, averaged waveforms of *s*-cut and *c*-cut samples are also shown.

### III.5. *d*-cut, *n*-cut and *g*-cut sapphire

Behavior of sapphire of *d*, *n*, and *g* orientations is similar to those described above. Figure 14 presents the VISAR waveforms measured for *d*-oriented sapphire samples of two different thicknesses at intermediate peak stress. Figure 15 summarizes results of experiments with 5-mm sapphire samples of *d*-orientation. The particle velocity histories are slightly less noisy as compared to *c*-cut sapphire and demonstrate lower Hugoniot elastic limit which also depends on the peak stress value. The HEL value practically does not depend on the peak stress. Even at smallest impact velocity the waveforms demonstrate strong relaxation whereas the peak stress corresponds to expected purely elastic response under these impact conditions.

Figure 16 summarizes results of experiments with 5-mm *n*-cut sapphire samples at three different peak stresses. *n*-cut sapphire demonstrates smallest HEL values over the whole range of stresses. A two-wave configuration is clearly recorded even at smallest impact velocity. The non-monotonic variation of the HEL with increasing peak stress is more pronounced than that of *r*-cut sapphire. Figure 17 presents results of experiments with *g*-cut sapphire. Its behavior may be considered as intermediate between *c*-cut and *r*-cut sapphire.

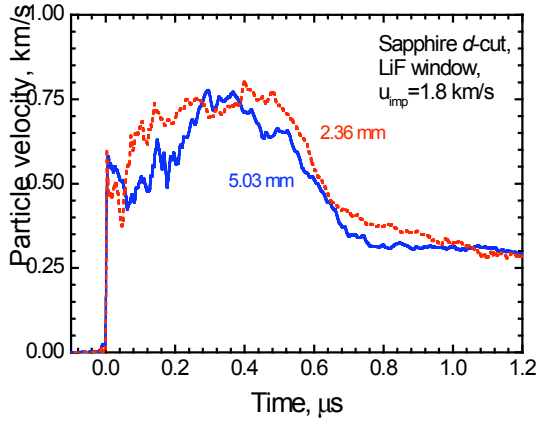
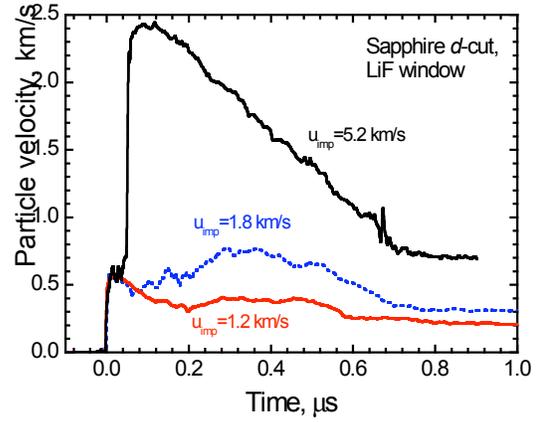

Fig. 14. Results of experiments with *d*-orientated sapphire samples 5.03 mm and 2.36 mm in thickness impacted by 2-mm aluminum flyer plate at 1.8 km/s of impact velocity through aluminum base plate 2 mm in thickness.

Fig. 15. Results of experiments with 5 mm-thick sapphire samples of *d*-orientation.

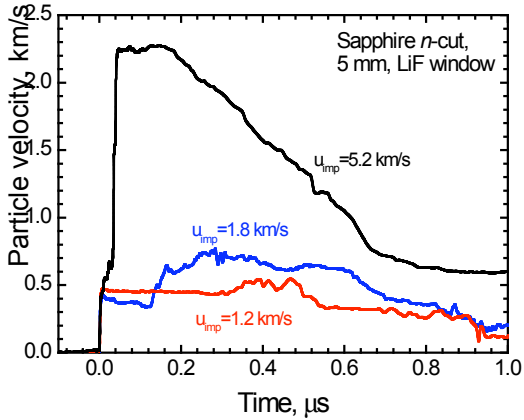
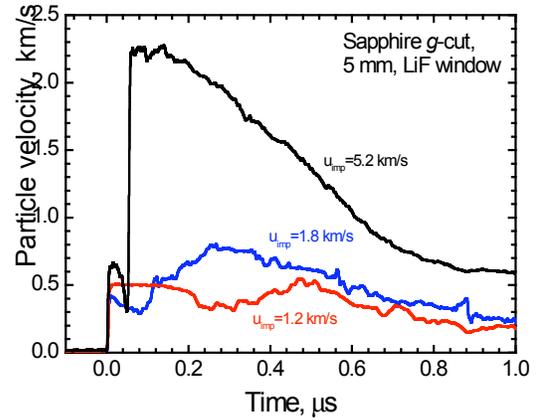

Fig. 16. Results of experiments with sapphire samples of *n*-orientation 5 mm in thickness at three different impact stresses.

Fig. 17. Results of experiments with sapphire samples of *g*-orientation.

**IV. Discussions**

Table 2 summarizes results of measurements of the HEL values for sapphire of seven tested orientations. The Hugoniot in the elastic compression range was obtained [1] for *c*-cut sapphire in the form $U_S = c_l + u_p$, where $U_S$ and $u_p$ are the shock front velocity and the particle velocity of shock compressed matter, respectively, and $c_l$ is the longitudinal sound speed at zero pressure. We used the same relationship for Hugoniots for sapphire of all other orientations with corresponding $c_l$ values. All data in the Table 2 are related to 5 mm-thick samples.

**Table 2**. Hugoniot elastic limits of sapphire (GPa).

| Impact velocity, km/s | Orientation | | | | | | |
|---|---|---|---|---|---|---|---|
| | $c$ (0001) | $d$ (10$\bar{1}$4) | $r$ (1$\bar{1}$02) | $n$ (11$\bar{2}$3) | $s$ (10$\bar{1}$1) | $g$ (11$\bar{2}$1) | $m$ (10$\bar{1}$0) |
| 1.2±0.05 | 16 | 17.3 | 13.3 | 14.1 | 15.9 | 15.7 | 15.1 |
| 1.8±0.05 | 17.7 | 16.7 | 12.8 | 12.4 | 13.6 | 12.8 | 18.3 |
| 5.2±0.1 | 24.2 | 19.7 | 18.7 | 12.4 | 20.9 | 20.5 | 22.3 |

These HEL values slightly exceed values reported by Graham and Brooks [4] and Mashimo et al. [8]. We see two possible reasons of this disagreement. Graham and Brooks made their measurements with thicker samples and obtained lower HEL values as a result of decay of the elastic-precursor waves. In both works [4] and [8] the diagnostic was based on recording the time-distance diagram of the movement of sample surface. This diagnostic does not allow resolution of fast velocity oscillations; it can be supposed these data are related rather to average stress between the elastic and plastic waves.

Irregular oscillations in measured waveforms are obviously associated with heterogeneity of deformation. These oscillations are strongest in both absolute and relative senses at low shock stresses and are smallest at high shock stresses. Accounting for small rise time (5-7 ns) of strong plastic shock waves we may suppose the heterogeneity of deformation decreases with increasing peak shock stress.

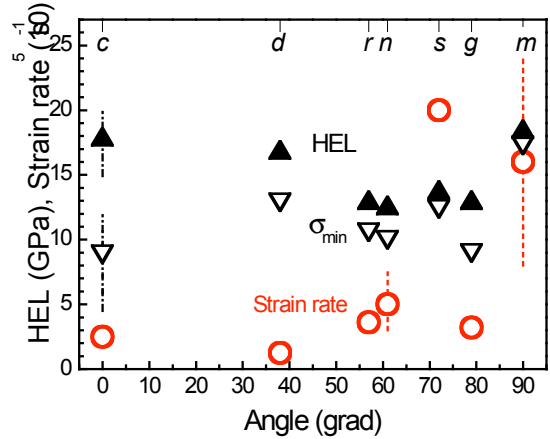

Fig. 18. Measured Hugoniot elastic limits (HEL), minimal stresses between elastic and plastic fronts $\sigma_{min}$, and strain rate at midpoint in time of plastic wave for sapphire as function of angle between direction of shock propagation normal and $c$ axis at 1.8 km/s impact velocity.

Figure 18 summarizes results of measurements of HELs and strain rates in plastic waves for the measured sapphire orientations at intermediate impact stress. Strain rate $\dot{\varepsilon}$ was estimated as $\dot{\varepsilon} \approx \dot{u}_{LiF} \left( \rho_{sap} c_{sap} + \rho_{LiF} c_{LiF} \right) / 2 \rho_{sap} c_{sap}^2$, where $\dot{u}_{LiF}$ is measured acceleration of the sapphire-LiF interface at the midpoint in time of the plastic-compression wave and $\rho_{sap}, c_{sap}, \rho_{LiF}, c_{LiF}$ are densities and bulk sound speeds of sapphire and lithium fluoride at 15-20 GPa pressure, respectively. These dependences are not monotonic and are not correlated to each other, which is a result of competition of different mechanisms of plastic deformation. A fast rise time means a high strain rate, which means the wave is the most similar to that of an ideal smooth plastic wave. The m-and s-cut orientations have the highest strain rates.

Possible mechanisms of plastic deformation which are discussed in literature include basal slip, pyramidal slip, prism slip, basal twinning, and rhombohedral twinning. The basal twins and dislocations have the lowest critical resolved shear stress to activation. Experimental data in Table 2 show the HEL values vary from 12.4 GPa up to 24.2 GPa depending on the load direction and the peak stress. Establishing a correlation between the measured HEL values and activated mechanisms of slip and twinning will be a subject of upcoming analysis. However, our initial view shows that highest HEL values are observed at shock compression perpendicular and parallel to the crystal base plane in experiments with *c*-cut and *m*-cut samples. The variation of longitudinal sound speed $c_l$ with angular orientation is similar to that of the HEL; that is, $c_l$ is largest along the c and m directions and least between these extremes. Uniaxial shock compression in the *c* and *m* directions excludes generation of shear stresses on the basal plane and, correspondingly, activation and movement of dislocations and twins in this plane. Shock compression along the *c*-axis also excludes prism slip. The shape of the waveforms recorded for *s*-cut and *m*-cut sapphire, as compared to data for other load directions, suggests with high probability that the highest contributions of dislocation mechanisms to plasticity are along these crystal directions.

Behavior of sapphire under shock compression is associated with a set of specific features, some of which have not been observed for other materials. Irregular oscillations of stress, which are most pronounced in the waveforms for *c*-cut sapphire, often accompany twinning. Similar oscillation, although not as strong as that for sapphire, were recorded on the waveforms in experiments with titanium [27], shock compression of which is accompanied by intense twinning. Twins can grow with extremely high speed – up to 8 km/s in sapphire [22]. The formation of each twin makes an essential contribution to the plastic strain and thereby causes stress relaxation. The unexpected and unusual decrease of recorded HEL values for all orientations, with the exception of m-cut samples (Fig. 12) with increasing impact velocity from 1.2 km/s to 1.8 km/s is probably explained by the fact that nucleation of twins in the crystal requires much higher stresses than their growth.

The low reproducibility of the waveforms points to stochastic nucleation of twins, like nucleation of bubbles in superheated liquids. If this analogy is true, the rate of nucleation of twins should grow exponentially with increasing applied stress. A higher density of twins provides more uniform distribution of plastic shears in the bulk of material, a higher strain rate and, correspondingly, deeper stress relaxation which appears at the intersection of the waveforms between elastic and plastic fronts. In turn, fast relaxation of stress behind the precursor front causes its hydrodynamic decay. Such a scenario can help understand the anomalous decrease of recorded HEL values with increasing peak shock stress but the question arises as to how the mechanisms of plastic deformation may be activated again at lowered stresses in the front of the second compression wave. Some contribution into nucleation of plastic shears is probably deposited by an interference of the stress field perturbations which have been engendered by preceding shears. One can hope that detailed analysis of the wave dynamics, anomalous dependences of recorded HEL values on the peak shock stress, and intersections of the waveforms in the stress relaxation region will help us to justify mechanisms and determine the kinetics of nucleation and growth of the deformation defects in sapphire.

Within the elastic deformations range sapphire demonstrates extremely high resistance to tensile fracture and its strong dependence on time. Nucleation of fractures has stochastic character and the trend for spall strength to decrease with increasing peak compressive stress indicates that the mean time of formation of a critical nucleation site perhaps depends on the duration of the stage of non-hydrostatic compression preceding the tension. This assumption implies that mechanisms of nucleation of fractures and of plastic shears involve common phenomena, such as twinning. Nucleation of cracks from the intersection of two twin systems was observed in molecular dynamic simulations [22] and explains the loss of strength of plastically deformed sapphire.

Concerning use of sapphire as a window material, the results of our experiments lead us to the following conclusions. Sapphire of *c*-orientation does not deform plastically and consequently remains homogeneous under shock compression up to 20-25 GPa for 0.1 μs or less. At shock compression above this range, heterogeneous inelastic deformation of *c*-cut sapphire destroys its optical uniformity. Average spacing between local shears (which characterize the degree of heterogeneity) decreases with increasing peak shock stress. Shock compression of *s*-cut and *m*-cut sapphires above their HELs is accompanied by the smallest heterogeneity of deformation. Obviously, the *m*-cut and *s*-cut orientations of sapphire are expected to be the best window materials.

This work was supported by the US Army Research Office through CRDF GAP grants number RUE2-1668-MO-06 and RUE2-31010-MO-08 and by the Program of Basic Research of Russian Academy of Sciences "Fundamental problems of mechanics of interactions in technical and natural systems, materials, and media". Work at HU was supported by U.S. Army Research Office Grant W911NF0610517. We acknowledge Princeton Scientific Corporation for providing the *c*-, *d*-, and *r*-cut sapphire crystals.